\documentclass[groupedaddress, superscriptaddress, citesort]{revtex4}
\usepackage{amsmath,amssymb,amsfonts,amsthm,fancyhdr}
\usepackage{graphicx}

\def\be{\begin{equation}}
\def\ee{\end{equation}}
\def\ba{\begin{array}}
\def\ea{\end{array}}

\begin{document}
\baselineskip=18pt


\title{INEQUALITIES DETECTING ENTANGLEMENT FOR ARBITRARY
BIPARTITE SYSTEMS}

\author{Hui Zhao}
\email{zhaohui@bjut.edu.cn}
\affiliation{College of Applied Sciences, Beijing University of
Technology, Beijing 100124, China}

\author{Shao-Ming Fei}
\email{feishm@cnu.edu.cn}
\affiliation{School of Mathematical Sciences, Capital Normal University,
Beijing 100048, China}

\author{Jiao Fan}
\email{fanjiao1015@126.com}
\affiliation{College of Applied Sciences, Beijing University of
Technology, Beijing 100124, China}

\author{Zhi-Xi Wang}
\email{wangzhx@cnu.edu.cn}
\affiliation{School of Mathematical Sciences, Capital Normal University,
Beijing 100048, China}

\begin{abstract}

 Based on the generators of $SU(n)$ we present
inequalities for detecting quantum entanglement for $2
\otimes d$ and $M \otimes N$ systems. These
inequalities provide a sufficient condition of entanglement
for bipartite mixed states and give rise to an experimental way
of entanglement detection.
\end{abstract}


\maketitle

\section{Introduction}	

Quantum entanglement has played very important roles in quantum
information processing such as quantum teleportation, quantum
cryptography, quantum dense coding and parallel computing
\cite{nielsen,horodecki,guehne}. One of the important problems in
the theory of quantum entanglement is to detect the quantum
entanglement by measuring some suitable quantum mechanical
observables. The Bell inequalities can be used to detect perfectly
the entanglement of pure bipartite states \cite{CHSH,N. Gisin92,D.
Collins,Chen7,Li8}. Besides Bell inequalities, the
entanglement witness are also useful in experimental detection of
quantum entanglement for mixed states \cite{M. Horodecki,Philipp10,Lewenstein11,wit2,wit3,wit4,wit5}.
For bipartite mixed states, a
necessary and sufficient inequality has been derived for detecting
entanglement of two-qubit states \cite{sixiayu}. The inequality in
\cite{Ming} is both necessary and sufficient in detecting
entanglement of qubit-qutrit states, and necessary for qubit-qudit
states. In Ref.\cite{Ming1} an inequality detecting entanglement for
arbitrary dimensional bipartite states has been presented.

In stead of particular construction of the quantum mechanical
observables in \cite{Ming,Ming1}, in this paper we use directly the
generators of $SU(n)$ and present new inequalities for detecting
entanglement of arbitrary dimensional bipartite mixed states. These
inequalities give a necessary condition of separability for general
mixed states. Any violation of the inequalities implies quantum
entanglement. The paper is organized in the following way. In Sec.
2, based on Pauli matrices and the generators of $SU(d)$, we
present inequalities for detecting entanglement of $2\otimes d$
systems. In Sec. 3, we present inequalities for detecting
entanglement of $M \otimes N$ systems. Conclusions are given in Sec.
4.

\section{Inequalities Detecting Entanglement for $2 \otimes d$ Systems}

Let $H$ be $n$-dimensional vector space with computational basis
$\{|i \rangle \}^{n}_{i=1}$. The generalized Gell-Mann matrices
(GGM) are the generators of $SU(n)$ defined by \cite{Bertlmann}:

\begin{itemize}
\item [(i)]$\frac{n(n-1)}{2}$ symmetric GGM
\begin{equation}
    \lambda^{jk}_{s} = |j \rangle \langle k| +|k \rangle \langle j|, \ \ 1\leq j < k\leq n;
\end{equation}
\item [(ii)] $\frac{n(n-1)}{2}$ antisymmetric GGM
\begin{equation}
    \lambda^{jk}_{\alpha} = -\texttt{i} |j \rangle \langle k| +\texttt{i}|k \rangle \langle j|, \ \ 1\leq j < k\leq n;
\end{equation}
\item [(iii)] $(n-1)$ diagonal GGM
\begin{equation}
    \lambda^{l} = \sqrt{\frac{2}{l(l+1)}} (\sum^{l}_{j=1}|j \rangle
    \langle j| - l|l+1 \rangle \langle l+1| ), \ \ 1\leq l \leq n-1.
\end{equation}
\end{itemize}

In total we have $n^{2}-1$ GGM which are Hermitian and traceless.
The operator $|j \rangle \langle k|$ with $j, k =1, \cdots, n$ can be also
expressed in terms of GGM \cite{Bertlmann}
\begin{equation}
|j \rangle \langle k| =\left\{
\begin{array}{ll}
\displaystyle\frac{1}{2} (\lambda^{jk}_{s} + \texttt{i} \lambda^{jk}_{\alpha}), & \hbox{for $j <k$;} \\[4mm]
             \displaystyle              \frac{1}{2} (\lambda^{kj}_{s} - \texttt{i} \lambda^{kj}_{\alpha}), & \hbox{for $j >k$;} \\[4mm]
              \displaystyle             - \sqrt{\frac{j-1}{2j}} \lambda^{j-1} + \sum^{n-j-1}_{m=0}\frac{1}{\sqrt{2(j+m)(j+m+1)}}
                           \lambda^{j+m} +\frac{1}{n}I, & \hbox{for j=k.}
                           \end{array}
                           \right.
\end{equation}

As any $2\times 2$ matrix can be expanded according to
the Pauli matrices plus identity, for the case of $n=2$, opertors $|j \rangle \langle k|$, $j,k =1, 2$, can be written as
\begin{eqnarray}
  |1 \rangle \langle 1| &=& \frac{1}{2} ( I_{2} + \sigma_{1}),~~~~~~~
  |2 \rangle \langle 2| = \frac{1}{2} ( I_{2} - \sigma_{1}), \nonumber \\
  |1 \rangle \langle 2| &=& \frac{1}{2} ( \sigma_{2} + \texttt{i} \sigma_{3}),~~~~~
  |2 \rangle \langle 1| = \frac{1}{2} ( \sigma_{2} - \texttt{i} \sigma_{3}),
\end{eqnarray}
where $\sigma_{i}$ $(i= 1,2,3)$ are the Pauli matrices. For the case
of $n=d$, $|j \rangle \langle k|$, $j, k =1, 2$, can be written as
\begin{eqnarray}
  |1 \rangle \langle 1| &=& \sum^{d-2}_{m=0}\frac{1}{\sqrt{2(m+1)(m+2)}}
  \lambda^{m+1} +\frac{1}{d}I, \nonumber \\
  |2 \rangle \langle 2| &=& -\frac{1}{2} \lambda^{1} +\sum ^{d-3}_{m=0}
  \frac{1}{\sqrt{2(m+2)(m+3)}} \lambda^{m+2} +\frac{1}{d}I, \nonumber \\
  |1 \rangle \langle 2| &=& \frac{1}{2} ( \lambda_{s}^{12} + \texttt{i} \lambda_{\alpha}^{12}), \nonumber \\
  |2 \rangle \langle 1| &=& \frac{1}{2} ( \lambda_{s}^{12} - \texttt{i} \lambda_{\alpha}^{12}).
\end{eqnarray}

Next we construct quantum mechanical operators for bipartite $2 \otimes d$ systems $A$ and $B$. Set
\begin{eqnarray}\label{Y123}
  \widehat{Y}_{1} &=& \frac{1}{2} ( \sigma_{2} + \texttt{i} \sigma_{3}) \otimes
  \frac{1}{2} ( \lambda_{s}^{12} - \texttt{i} \lambda_{\alpha}^{12}) +
  \frac{1}{2} ( \sigma_{2} - \texttt{i} \sigma_{3}) \otimes
  \frac{1}{2} ( \lambda_{s}^{12} + \texttt{i} \lambda_{\alpha}^{12}), \nonumber \\[4mm]
  \widehat{Y}_{2} &=&  \frac{1}{2} ( I_{2} + \sigma_{1}) \otimes
  (\sum^{d-2}_{m=0}\frac{1}{\sqrt{2(m+1)(m+2)}} \lambda^{m+1} +\frac{1}{d}I_{d} )  \nonumber \\
   & & -\frac{1}{2} ( I_{2} - \sigma_{1}) \otimes
   (-\frac{1}{2} \lambda^{1} +\sum^{d-3}_{m=0} \frac{1}{\sqrt{2(m+2)(m+3)}} \lambda^{m+2} +\frac{1}{d}I_{d}),  \nonumber\\[4mm]
  \widehat{Y}_{3} &=& \frac{1}{2} ( I_{2} + \sigma_{1}) \otimes
  ( \sum^{d-2}_{m=0}\frac{1}{\sqrt{2(m+1)(m+2)}}  \lambda^{m+1} +\frac{1}{d}I_{d} )  \nonumber \\
   & & + \frac{1}{2} ( I_{2} - \sigma_{1}) \otimes
   (-\frac{1}{2} \lambda^{1} +\sum^{d-3}_{m=0} \frac{1}{\sqrt{2(m+2)(m+3)}} \lambda^{m+2} +\frac{1}{d}I_{d}).
\end{eqnarray}
Denote $Y_{i}=Tr( \rho (U \otimes V ) \widehat{Y}_{i} ( U
\otimes V )^{\dag})$, $i=1, 2, 3$, where $U$ and $V$ are unitary
transformations on systems $A$ and $B$, respectively. We have the
following theorem:
\vspace*{12pt}

\noindent
{\bf Theorem 1:} \emph{Any separable state $\rho \in H_{2} \otimes H_{d}$
obeys the following inequality
\begin{equation}\label{T1}
Y^{2}_{3} \geq Y^{2}_{1} + Y^{2}_{2}.
\end{equation}}

\begin{proof}
First we prove that the inequality holds for product states. If $\rho$ is separable, its partial transposed
matrix $\rho^{T_{B}}$ is non-negative, i.e. $Tr(\rho^{T_{B}} P_{AB})
\geq 0$, where $P_{AB}$ is an arbitrary projector to $2 \otimes 2$
subsystems. Or more generally $Tr[\rho^{T_B}(U^{A}\otimes
U^{B})P_{AB}(U^{A}\otimes U^{B})^{\dag}]\geq 0$, where $U^{A}$ and
$U^{B}$ are local unitary operators. Any $2\otimes 2$ pure state has
the Schmidt decomposition,
\begin{equation}
|\phi \rangle = \sin \theta |1 \rangle_{A} \otimes |1 \rangle_{B} +
\cos \theta |2 \rangle_{A} \otimes |2 \rangle_{B}.
\end{equation}
Hence $P_{AB}$ can be written as
\begin{equation}
    P_{AB}=(U \otimes V) |\phi \rangle \langle \phi | (U \otimes V) ^{\dag},
\end{equation}
where $U$ and $V$ are unitary operators. We have
\begin{eqnarray}
  & &Tr[\rho^{T_B}(U^{A}\otimes U^{B})P_{AB}(U^{A}\otimes U^{B})^{\dag}] \\[2mm]
  &=& Tr \{ \rho [ (U^{A} \otimes U^{B})(U \otimes V)  |\phi \rangle \langle \phi |
   (U \otimes V) ^{\dag} (U^{A} \otimes U^{B})^{\dag}] ^{ T_{B}} \} \nonumber \\[2mm]
    &= & Tr \{ \rho  ( U^{A} U \otimes (U^{B} V)^{{\dag} { T_{B}}})  (|\phi \rangle
    \langle \phi |) ^{ T_{B}} (U^{A}U) ^{\dag} \otimes (U^{B} V)^{ T_{B}} \} \nonumber \\[2mm]
    &=&  Tr [ \rho (U \otimes V) (|\phi \rangle \langle \phi |) ^{ T_{B}}  (U \otimes V) ^{\dag} ],
\end{eqnarray}
where the last equation is obtained by chossing $U^{A} = I$ and $U^{B} = (V V ^{ T_{B}})^{\dag}$.
Therefore we get the following inequality,
\begin{equation}\label{ob}
Tr[\rho^{T_B}(U^{A}\otimes U^{B})P_{AB}(U^{A}\otimes
U^{B})^{\dag}]=Tr [ \rho (U \otimes V) (|\phi \rangle \langle \phi
|) ^{ T_{B}} (U \otimes V)^{\dag} ] \geq 0.
\end{equation}
Using Eqs. (5), (6), (7) and (9), we have
\begin{eqnarray}
  (|\phi \rangle \langle \phi |) ^{ T_{B}} &=&
  \sin ^{2} \theta |11 \rangle \langle 11| + \frac{1}{2} \sin 2\theta
  ( |12 \rangle \langle 21| + |21 \rangle \langle 12|)
  + \cos^{2} \theta |22 \rangle \langle 22| \nonumber \\
   &=& \frac{1}{2} (\widehat{Y}_{1} \sin 2\theta - \widehat{Y}_{2} \cos 2\theta
   +\widehat{Y}_{3} ).
\end{eqnarray}
Substituting Eq. (14) into inequality (13) and setting $t = \tan
\theta$, we obtain
\begin{equation}
 (Y_{2} + Y_{3})t^{2} + 2Y_{1} t + ( Y_{3}-Y_{2}) \geq 0.
\end{equation}
Since $ Y_{2} + Y_{3} = 2Tr(\rho(U\otimes V)|11 \rangle \langle
11|(U\otimes V)^\dag) \geq 0$ and (15) is valid for any $t$,
we get
\begin{equation}
Y^{2}_{1}+ Y^{2}_{2}-Y^{2}_{3} \leq 0.
\end{equation}
Therefore the inequality $ Y^{2}_{3} \geq Y^{2}_{1}+ Y^{2}_{2}$
holds for any product states.

It has been proved that if the
inequality $a^{2}_{i} \geq b^{2}_{i} + c^{2}_{i}$ holds for
arbitrary real numbers $b_{i}$ and $c_{i}$ and non-negative $a_{i}$,
$i =1, \cdots, n$, then
\begin{equation}
 (\sum^{n}_{i=1} p_{i} a_{i})^{2} \geq
 (\sum^{n}_{i=1} p_{i} b_{i})^{2} +
 (\sum^{n}_{i=1} p_{i} c_{i})^{2},
\end{equation}
for $0 \leq p_{i} \leq 1$ and $\sum^{n}_{i=1} p_{i} =1$ \cite{Ming}. For
general separable mixed states
\begin{equation}\label{ming}
\rho =\Sigma_{i} p_{i} |\phi_{i} \rangle \langle \phi_{i}|, \ \ 0
\leq p_{i} \leq 1, \ \  \sum_{i} p_{i}=1,
\end{equation}
where $|\phi_{i} \rangle $ are all product states, using (17)
one can verify that any mixed separable state $\rho$ also obeys the
inequality (\ref{T1}).
\end{proof}

{\noindent\bf Remark 1:} If $\rho$ is separable, the inequality $Tr[\rho^{T_B}(U^{A}\otimes
U^{B})P_{AB}(U^{A}\otimes U^{B})^{\dag}]\geq 0$ is valid for any local unitary operators
$U^{A}$ and $U^{B}$. In proving the theorem, we have chosen $U^A=I$ and $U^B={(VV^{T_B})}^\dagger$,
so that we can use the mean values of the set of quantum mechanical observables
$(U \otimes V ) \widehat{Y}_{i} ( U\otimes V )^{\dag}$ to detect quantum entanglement.

{\noindent\bf Remark 2:} The operators in (\ref{Y123}) are so introduced in terms of
the quantum mechanical observables: the Pauli matrices and the $SU(n)$ generators.
In fact, due to the direct relations between $|j \rangle \langle
k|$ and the generators of $SU(n)$, for bipartite quantum systems
$\widehat{Y_1}$, $\widehat{Y_2}$ and $\widehat{Y_3}$ can be simply written as
$\widehat{Y}_{1}=|12\rangle \langle 21|+|21\rangle \langle 12|$,
$\widehat{Y}_{2}=|11\rangle \langle 11|-|22\rangle \langle 22|$,
$\widehat{Y}_{3}=|11\rangle \langle 11|+|22\rangle \langle 22|$.

To show the advantage of our inequality, let us consider the following examples.
\noindent
{\bf Example 1:} The two-qubit Werner state is given by
\cite{Azuma},
\begin{equation}
W(a)=a | \Psi ^{-} \rangle  \langle  \Psi ^{-}  |  +
\frac{1-a}{4}I_{4}.
\end{equation}
where $0\leq a \leq 1$, $I_{4}$ denotes the $4\times 4$ identity
matrix. $| \Psi ^{-} \rangle $ is the maximally entangled two-qubit state,
\begin{equation}
| \Psi ^{-} \rangle = \frac{1}{\sqrt{2}}( | 1 \rangle _{A} | 2 \rangle _{B} -
| 2 \rangle_{A} | 1 \rangle _{B}).
\end{equation}
$W(a)$ is  separable for $0 \leq a \leq \frac{1}{3}$ and entangled for $\frac{1}{3}< a \leq 1$.

We may simply choose $U = I_{2}$
and $V = I_{2}$, then
$$
Y^{2}_{1}+Y^{2}_{2}-Y^{2}_{3}=\frac{1+a}{4}(3a -1)>0.
$$
Therefore $ W(a)$ is entangled for $\frac{1}{3}< a \leq 1$. Our
inequalities can detect all the entanglement. It is noted that the inequalities constructed in Ref.~\cite{Ming} can detect the
entanglement of $ W(a)$ only for $1 \leq a < \frac{-2+4\sqrt{5}}{19}\approx 0.37$ when the same $U$ and $V$ are used.

\noindent
{\bf Example 2:} Let us consider the $2\otimes 3$ mixed state \cite{Ming},
\begin{equation}
\rho =a | \Psi ^{+} \rangle  \langle  \Psi ^{+}  |  + \frac{1-a}{6}I_{6},
\end{equation}
where $| \Psi ^{+} \rangle = \frac{1}{\sqrt{2}}( | 11 \rangle+
| 22 \rangle).$ This state is entangled if and only if $a >\frac{1}{4}$.
If we choose $U = \cos p \,( | 1 \rangle \langle 1 | + | 2 \rangle \langle 2
|) + \sin p\, ( | 1 \rangle \langle 2 | - | 2 \rangle \langle 1 |)$
and $V = I_{3}$, then
$$
F\equiv Y^{2}_{1}+Y^{2}_{2}-Y^{2}_{3}=\frac{1+2a}{9}(6a \sin^2 p -2a-1).
$$
For $p=\frac{\pi}{2}$, we have $Y^{2}_{1}+ Y^{2}_{2}-Y^{2}_{3} >0
$. Therefore $ \rho$ is entangled for $a >\frac{1}{4}$. Our
inequalities can detect all the entanglement in $\rho$, see Fig. 1.

\begin{figure}[!htb]
\includegraphics[width=0.65\textwidth]{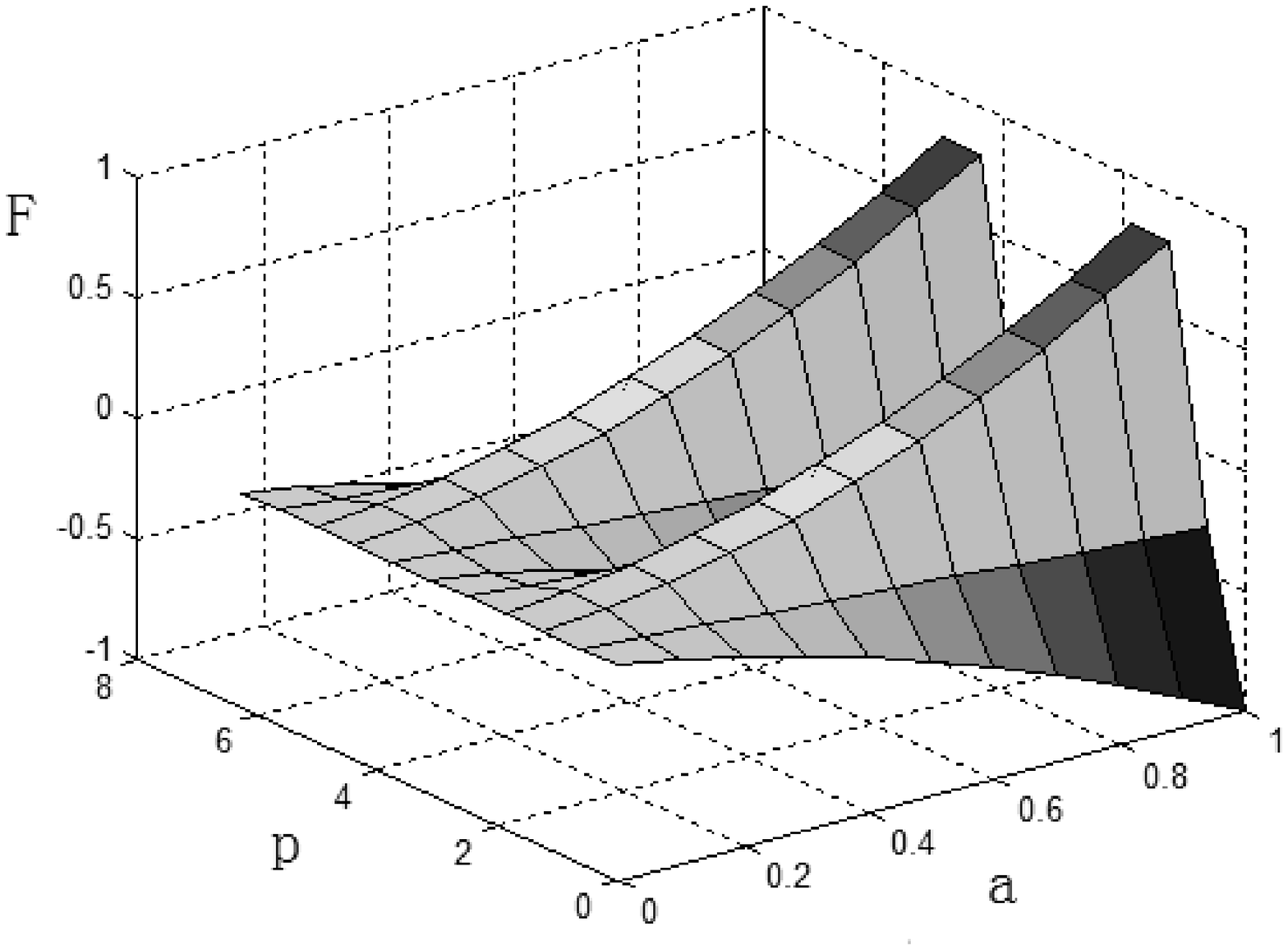}\includegraphics[width=0.65\textwidth]{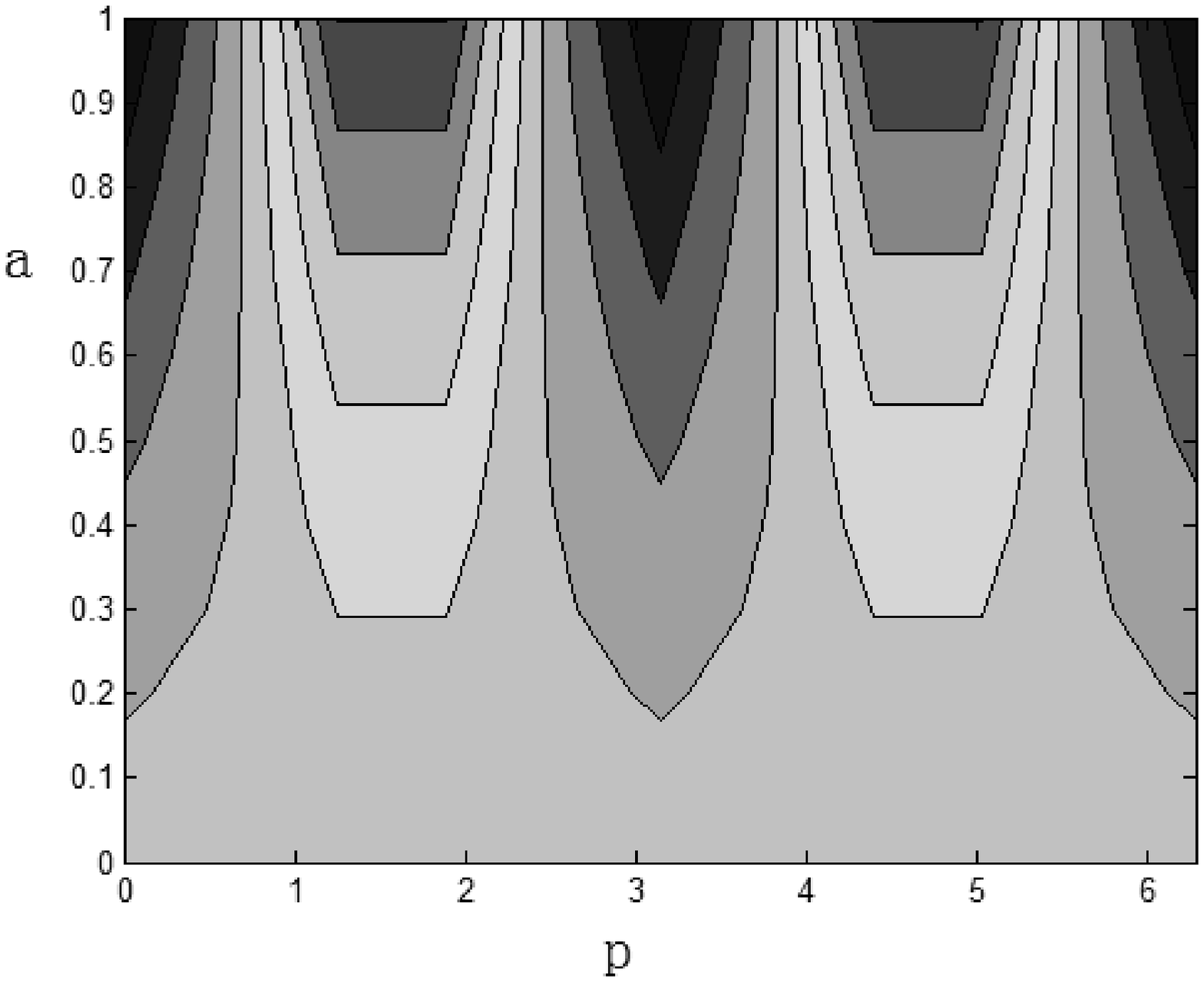}\\
\vspace*{8pt}
\caption{$U = \cos p ( | 1 \rangle \langle 1 | + | 2 \rangle \langle 2 |)
+ \sin p ( | 1 \rangle \langle 2 | - | 2 \rangle \langle 1 |)$, $V = I_{3}$.
Left figure: $F$ with respect to $p$ and $a$. Right figure: contour plot of the left figure.}
\end{figure}

\section{Inequalities Detecting Entanglement for $M \otimes N$ Systems}

Now we consider general $M\otimes N$ systems.
Let $H_{M}$, $H_{N}$ be $M$, $N$-dimensional vector spaces for systems $A$ and $B$,
respectively. For $j<k$, set
\begin{eqnarray}
  \widehat{Y}_{1} &=& \frac{1}{2} (\lambda^{jk}_{s} +\texttt{i} \lambda^{jk}_{\alpha})_{A}
  \otimes \frac{1}{2} (\lambda^{jk}_{s} - \texttt{i} \lambda^{jk}_{\alpha})_{B}
  +\frac{1}{2} (\lambda^{jk}_{s} - \texttt{i} \lambda^{jk}_{\alpha})_{A}
  \otimes \frac{1}{2} (\lambda^{jk}_{s} + \texttt{i} \lambda^{jk}_{\alpha})_{B}, \nonumber\\[4mm]
   \widehat{Y}_{2}  &=&(- \sqrt{\frac{j-1}{2j}} \lambda^{j-1} + \sum^{M-j-1}_{m=0}\frac{1}{\sqrt{2(j+m)(j+m+1)}}
   \lambda^{j+m} +\frac{1}{M}I) \otimes (- \sqrt{\frac{j-1}{2j}} \lambda^{j-1}\nonumber \\ & &+ \sum^{N-j-1}_{m=0}\frac{1}{\sqrt{2(j+m)(j+m+1)}}
   \lambda^{j+m} +\frac{1}{N}I) \nonumber \\
    & & -[(- \sqrt{\frac{k-1}{2k}} \lambda^{k-1} + \sum^{M-k-1}_{m=0}\frac{1}{\sqrt{2(k+m)(k+m+1)}}
   \lambda^{k+m} +\frac{1}{M}I) \otimes(- \sqrt{\frac{k-1}{2k}} \lambda^{k-1}\nonumber \\ & & + \sum^{N-k-1}_{m=0}\frac{1}{\sqrt{2(k+m)(k+m+1)}}
   \lambda^{k+m} +\frac{1}{N}I)], \nonumber\\[4mm]
  \widehat{Y}_{3}  &=&(- \sqrt{\frac{j-1}{2j}} \lambda^{j-1} + \sum^{M-j-1}_{m=0}\frac{1}{\sqrt{2(j+m)(j+m+1)}}
   \lambda^{j+m} +\frac{1}{M}I) \otimes (- \sqrt{\frac{j-1}{2j}} \lambda^{j-1}\nonumber \\ & &+ \sum^{N-j-1}_{m=0}\frac{1}{\sqrt{2(j+m)(j+m+1)}}
   \lambda^{j+m} +\frac{1}{N}I) \nonumber \\
    & &+(- \sqrt{\frac{k-1}{2k}} \lambda^{k-1} + \sum^{M-k-1}_{m=0}\frac{1}{\sqrt{2(k+m)(k+m+1)}}
   \lambda^{k+m} +\frac{1}{M}I) \otimes(- \sqrt{\frac{k-1}{2k}} \lambda^{k-1}\nonumber \\ & & + \sum^{N-k-1}_{m=0}\frac{1}{\sqrt{2(k+m)(k+m+1)}}
   \lambda^{k+m} +\frac{1}{N}I),
\end{eqnarray}
and $Y_{i}=Tr [\rho (U \otimes V) \widehat{Y}_{i} (U
\otimes V)^{\dag}]$ with $i=1, 2, 3,$ where $U$ and $V$ are local unitary
transformations on systems $A$ and $B$, respectively. We have the
following theorem:

\vspace*{12pt}
\noindent
{\bf Theorem 2:} \emph{Any separable state $\rho \in H_{M} \otimes H_{N}$
obeys the following inequality
\begin{equation}
Y^{2}_{3} \geq Y^{2}_{1} + Y^{2}_{2}.
\end{equation}}

\begin{proof}
Any product states can be written as
\begin{eqnarray}
|\xi \rangle =\sum^{M}_{i=1} \sum^{N}_{l=1} a_{i} b_{l} |il \rangle
,\end{eqnarray} with $\sum^{M}_{i=1} |a_{i}|^{2}= \sum^{N}_{l=1} |
b_{l}|^{2}=1 $. Using Eqs. (22) and (24), we have
\begin{eqnarray}
  Y^{2}_{3} - Y^{2}_{1} - Y^{2}_{2} &=& 4[ |a_{j} a_{k}b_{j} b_{k}|^{2}
  - Re^{2} (a_{j}a^{\ast}_{k} b_{j}^{\ast}b_{k})] \nonumber\\[2mm]
   &=& 4(a_{j} b_{k} a^{\ast}_{k} b^{\ast}_{j})  (a_{j} b_{k} a^{\ast}_{k} b^{\ast}_{j})^{\ast}
   - Re^{2} (a_{j} b_{k} a^{\ast}_{k} b^{\ast}_{j}) \geq  0.
\end{eqnarray}
Therefore $Y^{2}_{3} \geq Y^{2}_{1} + Y^{2}_{2}$ holds for any
product states. Using the similar methods in proving Theorem 1, we have that the
inequality also holds for general separable mixed states.
\end{proof}

To show the usefulness of our inequality, let us consider the Horodecki's \(3\,\otimes \,3\) state:
\begin{equation}\label{33}
\begin{aligned} \sigma _\alpha = \frac{2}{7}|\psi ^+\rangle \langle \psi ^+| +\frac{\alpha }{7}\sigma _+ +\frac{5-\alpha }{7}\sigma _-, \end{aligned}
\end{equation}
where $\sigma _+=\frac{1}{3}(|12\rangle \langle 12| + |23\rangle \langle 23| + |31\rangle \langle 31|)$,
$\sigma _-=\frac{1}{3}(|21\rangle \langle 21| + |32\rangle \langle 32| + |13\rangle \langle 13|)$,
$|\psi ^+\rangle =\frac{1}{\sqrt{3}}(|11\rangle + |22\rangle + |33\rangle )$. $\sigma _\alpha$
is separable for \(2\le \alpha \le 3\), bound entangled for \(3< \alpha \le 4\),
and free entangled for \(4< \alpha \le 5\) \cite{HorodeckiP}.
If we choose $U = \cos p \,( | 1 \rangle \langle 1 | + | 2 \rangle \langle 2
|) + \sin p\, ( | 1 \rangle \langle 2 | - | 2 \rangle \langle 1 |)$
and $V = I_{3}$, then
$$
F=Y^{2}_{1}+Y^{2}_{2}-Y^{2}_{3}=(\frac{2}{21})^2[(\alpha^2-5
\alpha+10) \sin^4 p -2\sin^2 p-4].
$$
For $p=\frac{\pi}{2}$, we have $Y^{2}_{1}+ Y^{2}_{2}-Y^{2}_{3} >0 $.
Therefore $ \sigma _\alpha$ is entangled for $\alpha >4$.

In Ref.\cite{Ming1} an inequality for detecting entanglement of
arbitrary dimensional bipartite systems has been presented, which
can also detect the entanglement of (\ref{33}) for $\alpha >4$.
The quantum mechanical observables in our inequalities are
constructed systematically according to $SU(n)$ generators,
in contract to artificial constructions of observable operators in Ref.\cite{Ming1}. The mean values can be easily
calculated. In fact, due to the direct relations between $|j \rangle \langle
k|$ and the generators of $SU(n)$, for bipartite quantum systems
$\widehat{Y_1}$, $\widehat{Y_2}$ and $\widehat{Y_3}$ can be simply written as
$\widehat{Y}_{1}=|jk\rangle \langle kj|+|kj\rangle \langle jk|$,
$\widehat{Y}_{2}=|jj\rangle \langle jj|-|kk\rangle \langle kk|$,
$\widehat{Y}_{3}=|jj\rangle \langle jj|+|kk\rangle \langle kk|$,
where $j=1, 2, \ldots, M$, $k=1, 2, \ldots, N$ and $j<k$.

\section{Conclusions}

We have presented inequalities for detecting quantum entanglement of $2 \otimes d$
and $M \otimes N$ systems. These inequalities give necessary conditions of separability for mixed states.
Since these inequalities are given by quantum mechanical observables, namely, Hermitian
operators, they supply experimental ways of detecting entanglement by measuring
the mean values of these local observables.
For example, the state defined in Eq. (21) violates the inequalities for
quantum mechanical observables given by choosing
$U = | 1 \rangle \langle 2 | - | 2 \rangle \langle 1 |$ and $V=I$. Any
violation of the inequalities implies that the quantum states are entangled.
As for examples, it has been shown that our inequalities can detect entanglement well
when the measurement operators are suitably chosen.
Our inequalities are complemental to the existing ones. As the inequalities are directly
given by the $SU(n)$ generators acting on local subsystems, the approach can be readily
generalized to deal with the detection of entanglement for multipartite systems.

\section*{Acknowledgments}

This work is supported
by the National Natural Science Foundation of China (11101017
and 11275131), Beijing Natural Science Foundation Program and Scientific
Research Key Program of Beijing Municipal Commission of Education
(KZ201210028032) and the Importation and Development of High-Caliber Talents Project of Beijing Municipal Institutions (CITTCD201404067).

.

\end{document}